\documentclass[amsmath, amsfonts, showpacs, twocolumn, prb]{revtex4}
\usepackage{graphicx}
\usepackage{epsfig}
\usepackage{bm}
\usepackage{amsmath}
\usepackage{amssymb}
\usepackage{euscript}

\def\tphi{\tau_{\phi}}
\def\ts{\tau_{s}}
\def\tgl{\tau_{\mathrm{GL}}}

\begin{document}

\title{Transport theory of superconductors with singular interaction corrections}

\author{Alex Levchenko}
\affiliation{Materials Science Division, Argonne National
Laboratory, Argonne, Illinois 60439, USA}

\begin{abstract}
We study effects of strong fluctuations on the transport properties
of superconductors near the classical critical point. In this regime
conductivity is set by the delicate interplay of two competing
effects. The first is that strong electron-electron interactions in
the Cooper channel increase the life-time of fluctuation Cooper
pairs and thus enhance conductivity. On the other hand, quantum
pair-breaking effects tend to suppress superconductivity. An
interplay between these processes defines new regime,
$Gi\lesssim\frac{T-T_c}{T_c}\lesssim\sqrt{Gi}$, where fluctuation
induced transport becomes more singular, here $Gi$ is the Ginzburg
number. The most singular contributions to the conductivity stem
from the dynamic Aslamazov-Larkin term, and novel Maki-Thompson and
interference corrections. The crossover temperature $T_c\sqrt{Gi}$
from weakly to strongly fluctuating regime is generated
self-consistently as the result of scattering on dynamic variations
of the order parameter. We suggest that the way to probe
nonlinear-fluctuations in superconductors is by magnetoconductivity
measurements in the perpendicular field.
\end{abstract}

\date{April 30, 2010}

\pacs{74.25.F-, 74.40.-n}

\maketitle

\begin{figure}[b!]
\begin{center}\includegraphics[width=8cm]{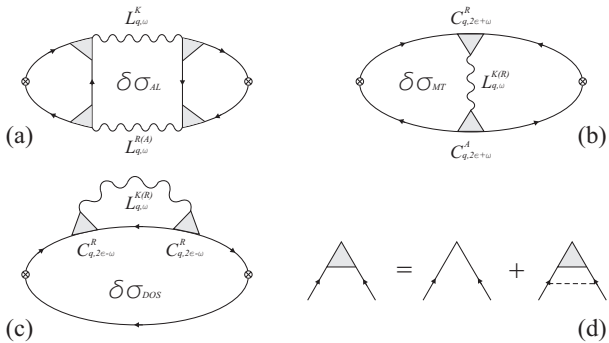}\end{center}
\caption{Superconductive fluctuation corrections to the normal-state
Drude conductivity: (a) Aslamazov-Larkin diagram, (b) Maki-Thompson
correction, (c) density of states contribution, and (d) Cooperon
impurity vertex. \label{Fig-1}}
\end{figure}

\textit{Introduction}. In the context of transport properties of
disordered fluctuating superconductors one usually discusses three
types of contributions to the normal-state Drude conductivity
$\sigma_D=e^2\nu D$ near $T_c$, see
Ref.~\onlinecite{Larkin-Varlamov}, here $D$ is the diffusion
coefficient and $\nu$ is the single-particle density of states. The
first one, $\delta\sigma_{AL}$, is called Aslamazov-Larkin (AL)
contribution.~\cite{Aslamazov-Larkin} It has a simple physical
interpretation as the direct charge transfer mediated by the
fluctuation Cooper pairs. Within the microscopic formulation, see
corresponding diagram in the Fig.~\ref{Fig-1}a, it reads
analytically
\begin{equation}\label{sigma-AL}
\frac{\delta\sigma_{AL}}{\sigma_{D}}=\frac{\pi}{32\nu
T^{2}_{c}}\sum_{q}Dq^2
\int\frac{d\omega}{\sinh^{2}\frac{\omega}{2T}}
\big[\mathrm{Im}L^{R}_{q,\omega}\big]^2\,,
\end{equation}
where
$L^{R}_{q,\omega}=-\frac{8T}{\pi}[Dq^2+\tgl^{-1}-i\omega]^{-1}$ is
the retarded component of the interaction (fluctuation) propagator,
and
$\tgl^{-1}=\frac{8T}{\pi}\ln\frac{T}{T_c}\backsimeq\frac{8}{\pi}(T-T_c)$
is inverse Ginzburg-Landau time. After the energy and momentum
integrations Eq.~\eqref{sigma-AL} reduces to the celebrated
result~\cite{Aslamazov-Larkin}
\begin{equation}\label{AL}
\frac{\delta\sigma_{AL}}{\sigma_{D}}=\frac{1}{2\pi
g}(T_c\tgl)=Gi\left(\frac{T_c}{T-T_c}\right)\,,
\end{equation}
for the thin-film superconductors with the thickness
$b\lesssim\sqrt{D\tgl}$, where $g=1/\nu Db$ is the dimensionless
conductance and $Gi=1/16g$ is the Ginzburg number. It is worth
recalling that AL term can be calculated from the time-dependent
Ginzburg-Landau theory and to some extent is classical.

The other two contributions have purely quantum origin. The
Maki-Thompson (MT) correction to conductivity,~\cite{Maki-Thompson}
$\delta\sigma_{MT}$, can be understood as the coherent Andreev
reflection of electrons on the local fluctuations of the order
parameter. Its most singular part near $T_c$ is given by
\begin{equation}\label{sigma-MT}
\frac{\delta\sigma_{MT}}{\sigma_{D}}=-\frac{1}{2\pi\nu T}\sum_q\int
\frac{d\epsilon
d\omega\coth\frac{\omega}{2T}}{\cosh^2\frac{\epsilon}{2T}}
\big[\mathrm{Im}
L^{R}_{q,\omega}\big]\big|C^R_{q,2\epsilon+\omega}\big|^2,
\end{equation}
which is shown diagrammatically in the Fig.~\ref{Fig-1}b. Here
$C^{R}_{q,\epsilon}=[Dq^2+\ts^{-1}-i\epsilon]^{-1}$ is the Cooperon,
which accounts for the scattering by impurities in the
particle-particle channel [sum of the ladder-type diagrams in the
Fig.~\ref{Fig-1}d], and $\ts$ is the spin-flip time. After the
integrations in 2d-case Eq.~\eqref{sigma-MT} reduces
to~\cite{Maki-Thompson}
\begin{equation}\label{MT}
\frac{\delta\sigma_{MT}}{\sigma_{D}}=\frac{1}{\pi
g}\frac{T_c\tgl}{1-\tgl/\ts}\ln\left(\frac{\ts}{\tgl}\right)\,,
\end{equation}
which unlike AL term [Eq.~\eqref{AL}] exhibits strong sensitivity to
the dephasing time and is formally divergent without pair-breaking
processes (no magnetic impurities for example). This is famous
feature of the MT correction.

Finally, the density of states (DOS) effects~\cite{Dos} originate
from the depletion of the energy states near the Fermi level by
superconductive fluctuations. It leads to the correction to
conductivity of the form [see Fig.~\ref{Fig-1}c]
\begin{eqnarray}\label{sigma-DOS}
\frac{\delta\sigma_{DOS}}{\sigma_D}=\frac{1}{2\pi\nu T} \sum_q\int
\frac{d\epsilon
d\omega\coth\frac{\omega}{2T}}{\cosh^2\frac{\epsilon}{2T}}\nonumber\\
\big[\mathrm{Im}
L^{R}_{q,\omega}\big]\mathrm{Re}\big(C^R_{q,2\epsilon-\omega}\big)^2\,,
\end{eqnarray}
which in contrast to AL and MT contributions is negative but has
much weaker (logarithmic instead of the power-law) temperature
dependence
\begin{equation}\label{DOS}
\frac{\delta\sigma_{DOS}}{\sigma_{D}}=-\frac{7\zeta(3)}{\pi^4g}\ln(T_{c}\tgl)\,,
\end{equation}
where $\zeta(x)$ is the Riemann zeta function.

Applicability of the perturbative treatment for superconductive
fluctuations implies that corresponding corrections to the
conductivity [Eqs.~\eqref{AL}, \eqref{MT}, and \eqref{DOS}] are
small as compared to its bare Drude value. Thus, requirement that
$\delta\sigma_{AL}+\delta\sigma_{MT}\lesssim \sigma_D$ restricts
perturbation theory to the temperatures above the Ginzburg region,
$T_cGi\lesssim T-T_c$. However, as it has been shown by Larkin and
Ovchinnikov,~\cite{Larkin-Ovchinnikov} this conclusion is premature.
It turns out that Eq.~\eqref{AL} is applicable only as long as
$T-T_c\gtrsim T_c\sqrt{Gi}$ while $\delta\sigma_{AL}$ becomes more
singular in the immediate vicinity of the critical temperature
$Gi\lesssim \frac{T-T_c}{T_c}\lesssim \sqrt{Gi}$
where~\cite{Larkin-Ovchinnikov}
\begin{equation}\label{AL-LO}
\frac{\delta\sigma_{AL}}{\sigma_{D}}\simeq\frac{1}{\pi^3g^2}
(T_c\tgl)^2(T_c\tphi),\,\,\,\,
\tphi^{-1}=\mathrm{max}\{\ts^{-1},T\sqrt{Gi}\}\,.
\end{equation}
In addition it was demonstrated by Reizer~\cite{Reizer} that MT term
saturates near $T_c$
\begin{equation}\label{MT-Reizer}
\frac{\delta\sigma_{MT}}{\sigma_{D}}\simeq
\frac{1}{2\pi\sqrt{g}}\ln\left(\frac{T_c\tgl}{\sqrt{g}}\right)\,,\quad
Gi\lesssim\frac{T-T_c}{T_c}\lesssim \sqrt{Gi},
\end{equation}
even without an extrinsic pair-breaking, such as magnetic
impurities. Thus, it was concluded that interaction corrections to
the conductivity of fluctuating superconductors are governed by
$\delta\sigma_{AL}$ [Eq.~\eqref{AL-LO}] at
$Gi\lesssim\frac{T-T_c}{T_c}\lesssim \sqrt{Gi}$. Stronger
singularity of the AL term was attributed to the life-time
enhancement of the preformed Cooper pairs by nonlinear-fluctuation
effects.~\cite{Larkin-Ovchinnikov} At the same time, saturation of
the interference MT contribution emerges as the result of scattering
on dynamical variations of the order parameter, which generate an
intrinsic dephasing time $\tphi^{-1}\simeq T\sqrt{Gi}\simeq
T/\sqrt{g}$.~\cite{Reizer,Brenig} In what follows we show that this
physical picture of fluctuations-enhanced transport is incomplete.
We identify novel class of interaction corrections, not discussed in
the literature before, which strongly influence conductivity at the
onset of superconducting transition. At the technical level
diagrammatic analysis for the conductivity corrections is carried
within nonlinear sigma model of fluctuating superconductors, see
Ref.~\onlinecite{Review} for the review.

\begin{figure}[t!]
\begin{center}\includegraphics[width=8cm]{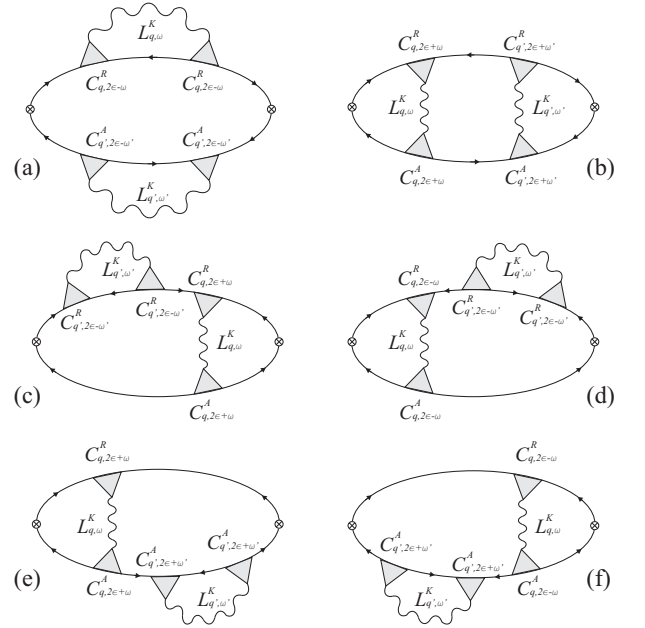}\end{center}
\caption{Singular corrections to the fluctuation conductivity:
diagrams (a) and (b) represent Maki-Thompson contributions due to
interaction of fluctuations and their interference correspondingly;
diagrams (c)-(f) represent mixture of the density of states and
Maki-Thompson scattering processes.\label{Fig-2}}
\end{figure}

\textit{Singular interaction corrections}. Apart from the
conventional contributions to the conductivity, which appear to the
first order in superconductive fluctuations, there are certain next
leading order terms which make conductivity to be more singular near
the transition. We find that among those the contributions shown in
the Fig.~\ref{Fig-2} are the most important. Obviously, these terms
carry and extra small pre-factor, $Gi\ll1$, due to the perturbative
treatment of fluctuation effects, however, they exhibit much
stronger temperature dependence then $\delta\sigma_{AL}$ and
$\delta\sigma_{MT}$, and become more important in the temperature
region $Gi\lesssim\frac{T-T_c}{T_c}\lesssim \sqrt{Gi}$.

We find two new MT contributions which depend differently on the
dephasing time. The first one is presented in the Fig.~\ref{Fig-2}a
and its most singular part near $T_c$ reads analytically as
\begin{eqnarray}
\frac{\delta\sigma^{a}_{MT}}{\sigma_D}=-\frac{\pi}{\nu^2}\sum_{qq'}
\int\frac{d\omega
d\omega'd\epsilon}{\cosh^2\frac{\epsilon}{2T}}L^{K}_{q,\omega}L^{K}_{q',\omega'}\nonumber\\
\times(C^R_{q,2\epsilon-\omega})^2(C^A_{q',2\epsilon-\omega'})^2\,.\label{sigma-MT-a}
\end{eqnarray}
This term represents interaction of superconductive fluctuations.
Notice here, that although diagrammatically $\delta\sigma^{a}_{MT}$
looks like second-order DOS effect, in fact, it should be classified
as MT term by the analytical properties. Indeed, MT contributions
involve mixture of retarded and advanced Cooperons while DOS terms
always bring Cooperons of the same causality. This important
difference makes DOS contributions to be subleading in powers of
$T_c\tgl$ [compare Eqs.~\eqref{sigma-MT} and \eqref{MT} with
Eqs.~\eqref{sigma-DOS} and\eqref{DOS}]. It is worth recalling that
$\delta\sigma^{a}_{MT}$ is already familiar from the studies of
diffusive~\cite{Varlamov-Dorin} and
ballistic~\cite{Ballistic-Tunnel} tunnel junctions, and granular
superconductors~\cite{Granular} above $T_c$. Assuming static
pair-breaking at this stage and after the consecutive energy
integrations Eq.~\eqref{sigma-MT-a} reduces to
\begin{eqnarray}
\frac{\delta\sigma^{a}_{MT}}{\sigma_D}=\frac{16T^{3}_{c}}{\pi\nu^2}\sum_{qq'}\frac{1}{(Dq^2+\tgl^{-1})
(Dq'^2+\tgl^{-1})}\nonumber\\
\times\frac{1}{(DQ^2+\mathrm{max}\{\ts^{-1},\tgl^{-1}\})^3}\,,\label{qq-sigma-MT-a}
\end{eqnarray}
where $Q^2=q^2+q'^2$. The remaining $q$ sums depend significantly on
the effective dimensionality. For the quasi-two-dimensional case we
find
\begin{equation}\label{MT-a}
\frac{\delta\sigma^{a}_{MT}}{\sigma_{D}}=\frac{1}{\pi^3g^2}\left\{
\begin{array}{lc}
\frac{\pi^2-9}{6}(T_c\tgl)^3 & \ts\gg\tgl,\\
\frac{}{} \\ (T_c\ts)^3\ln^{2}(\tgl/\ts) & \ts\ll\tgl\,.
\end{array}
\right.
\end{equation}
One special feature of this result is that it remains finite even in
the absence of extrinsic phase breaking, when $\ts\to\infty$. This
is unlike the other MT contribution shown in Fig.~\ref{Fig-2}b,
which represents an interference of superconductive fluctuations and
its most singular part is given by
\begin{eqnarray}
\frac{\delta\sigma^{b}_{MT}}{\sigma_{D}}=
-\frac{\pi}{\nu^2}\sum_{qq'} \int\frac{d\omega
d\omega'd\epsilon}{\cosh^2\frac{\epsilon+\omega'}{2T}}L^{K}_{q,\omega}L^{K}_{q',\omega'}\nonumber\\
\times|C^R_{q,2\epsilon+\omega}|^2|C^A_{q',2\epsilon+\omega'}|^2.\label{sigma-MT-b}
\end{eqnarray}
Technically, it is different from Eq.~\eqref{sigma-MT-a} by the
structure of the Cooperon propagators. Notice that the absolute
value of $C^{R(A)}$ makes the integrand of Eq.~\eqref{sigma-MT-b} to
be extended in the energy space whereas corresponding expression in
Eq.~\eqref{sigma-MT-a} is short ranged due to the pole structure of
the $(C^R)^2(C^A)^2$ product. This feature translates into the
different temperature dependence of $\delta\sigma^{b}_{MT}$ than
that of $\delta\sigma^{a}_{MT}$. Indeed, after energy integrations
Eq.~\eqref{sigma-MT-b} can be brought to the form
\begin{eqnarray}\label{qq-sigma-MT-b}
&&\hskip-1cm\frac{\delta\sigma^{b}_{MT}}{\sigma_{D}}=\frac{16T^{3}_{c}}{\pi\nu^2}
\sum_{qq'}\frac{1}{(Dq^2+\tgl^{-1})(Dq'^2+\tgl^{-1})}\nonumber\\
&&\hskip-1cm\frac{1}{(Dq^2+\ts^{-1})(Dq'^2+\ts^{-1})(DQ^2+\mathrm{max}\{\ts^{-1},\tgl^{-1}\})}\,,
\end{eqnarray}
which gives eventually in two dimensions with the logarithmic
accuracy
\begin{equation}\label{MT-b}
\frac{\delta\sigma^{b}_{MT}}{\sigma_{D}}\simeq\frac{1}{\pi^3g^2}\left\{
\begin{array}{lc}
(T_c\tgl)^3\ln^{2}(\ts/\tgl) & \ts\gg\tgl,\\
\frac{}{} \\
(T_c\ts)^3\ln^{2}(\tgl/\ts) & \ts\ll\tgl\,.
\end{array}
\right.
\end{equation}
Clearly, when compared to the corresponding limit of
Eq.~\eqref{MT-a}, the divergence of $\delta\sigma^{b}_{MT}$ at weak
pair-breaking $\ts\to\infty$, is the manifestation of coherence.

The remaining four terms in Figs.~\ref{Fig-2}c-\ref{Fig-2}f
represent the mixture of MT and DOS contributions, we thus label
those collectively by $\delta\sigma_{MTD}$. Although each individual
diagram has slightly different analytical structure, their most
divergent parts, however, are the same for all four terms. We find
then for the sum of these contributions,
$\delta\sigma_{MTD}\simeq4\delta\sigma_{\mathrm{Fig}-2c}$, following
expression
\begin{eqnarray}
\frac{\delta\sigma_{MTD}}{\sigma_D}=\frac{4\pi}{3\nu^2}\sum_{qq'}
\int\frac{d\omega
d\omega'd\epsilon}{\cosh^2\frac{\epsilon+\omega}{2T}}L^{K}_{q,\omega}L^{K}_{q',\omega'}\nonumber\\
\times|C^R_{q,2\epsilon+\omega}|^2(C^R_{q',2\epsilon-\omega'})^2\,.\label{sigma-MTD}
\end{eqnarray}
After the standard steps of integration this equation reduces to
\begin{eqnarray}\label{qq-sigma-MTD}
&&\hskip-1cm\frac{\delta\sigma_{MTD}}{\sigma_{D}}=-\frac{32T^{3}_{c}}{3\pi\nu^2}
\sum_{qq'}\frac{1}{(Dq^2+\tgl^{-1})(Dq'^2+\tgl^{-1})}\nonumber\\
&&\hskip-1cm\frac{1}{(Dq^2+\ts^{-1})(DQ^2+\mathrm{max}\{\ts^{-1},\tgl^{-1}\})^2}\,,
\end{eqnarray}
which gives for the conductivity correction of a thin
superconducting film
\begin{equation}\label{MTD}
\frac{\delta\sigma_{MTD}}{\sigma_{D}}\simeq-\frac{1}{3\pi^3g^2}\left\{
\begin{array}{lc}
(T_c\tgl)^3\ln(\ts/\tgl) & \ts\gg\tgl,\\
\frac{}{} \\
2(T_c\ts)^3\ln^{2}(\tgl/\ts) & \ts\ll\tgl\,.
\end{array}
\right.
\end{equation}
We see from here that mixed contributions suppress fluctuation
conductivity, unlike MT terms. Second, mixed terms exhibit weaker
divergence then $\delta\sigma^{b}_{MT}$ for the small static
pair-breaking $\ts\to\infty$. These singular MT and mixed
contributions [Eqs.~\eqref{MT-a}, \eqref{MT-b}, and \eqref{MTD}]
together with the AL correction Eq.~\eqref{AL-LO} represent the
leading terms in the asymptotic expansion of fluctuation
conductivity at the onset of superconducting transition.~\cite{note}

\begin{figure}[t!]
  \includegraphics[width=8cm]{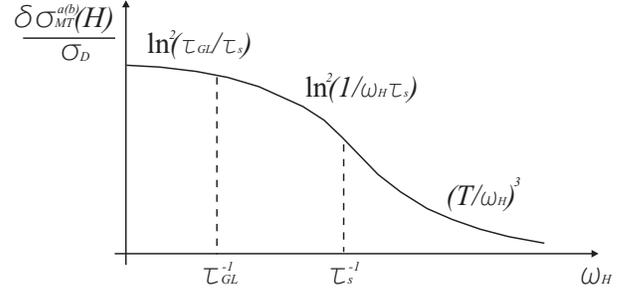}\\
  \caption{Sketch for the characteristic magnetic field dependence of the singular Maki-Thompson correction to
  the conductivity [Eqs.~\eqref{sigma-MT-a-H}]
  under the condition of strong phase-breaking scattering $\tgl\gg\ts$. In the opposite case
  one should replace $\tgl\rightleftarrows\ts$.}\label{Fig-3}
\end{figure}

\textit{Magnetoconductivity}. Since $\delta\sigma^{a(b)}_{MT}$ and
$\delta\sigma_{MTD}$ have distinct temperature dependence there is a
way to identify these terms by the appropriate transport experiment.
The most suitable one would be the magneto-conductivity measurement
in the perpendicular field. Indeed, magnetic field acts as an
effective pair-breaking mechanism which drives a superconductor away
from the critical region. This is simply understood by looking at
the pole structure of the interaction propagator
$L^{R(A)}(q,\omega)\propto(Dq^2+\tgl^{-1}\mp i\omega)^{-1}$ and
recalling that magnetic field $H$ applied perpendicularly to a film
changes the continuous spectrum of superconducting fluctuations into
its quantized form $Dq^2\to\omega_n=\omega_{H}(n+1/2)$, where
$\omega_{H}=4eDH$ is the cyclotron frequency and $n=0,1,2,\ldots$ is
the number of the Landau level. It becomes clear now that if
cyclotron frequency $\omega_H$ exceeds $\tgl^{-1}$, it is $\omega_H$
that cuts all energy transfer integrations, since
$\omega\sim\mathrm{max}\{\omega_{H},\tgl^{-1}\}$. Roughly speaking
it means that in the expressions for the conductivity corrections
Eqs.~\eqref{MT-a}, \eqref{MT-b} and \eqref{MTD} the scale of $T-T_c$
is replaced by $\omega_H$. This gives a possibility to restore the
temperature dependence of $\delta\sigma(T)$ by observing its
behavior as the function of magnetic $H$, which traces corresponding
dependence on $T-T_c$. In what follows we calculate
$\delta\sigma^{a}_{MT}(H)$ explicitly and quote only final results
for the remaining contributions.

Starting from Eq.~\eqref{qq-sigma-MT-a} we replace momentum
integration by the discrete sum over the Landau levels
$\sum_q\to\frac{\omega_H}{4\pi Db}\sum^{\infty}_{n=0}$, where the
prefactor accounts for the degeneracy in the position of the orbit,
and find
\begin{eqnarray}
\frac{\delta\sigma^{a}_{MT}(H)}{\sigma_D}=\frac{\omega^2_HT^3_{c}}{\pi^3g^2}\sum^{\infty}_{nn'=0}
\frac{1}{(\omega_{n}+\tgl^{-1})
(\omega_{n'}+\tgl^{-1})}\nonumber\\
\frac{1}{(\Omega_{nn'}+\mathrm{max}\{\ts^{-1},\tgl^{-1}\})^3},\label{sigma-MT-a-H}
\end{eqnarray}
where $\Omega_{nn'}=\omega_n+\omega_{n'}$. If
$\omega_H\ll\mathrm{max}\{\tgl^{-1},\ts^{-1}\}$, which corresponds
to the zero-field limit, we restore Eq.~\eqref{MT-a}. At higher
fields there are two regimes. For $\tgl^{-1}\ll\omega_H\ll\ts^{-1}$
quantization of the spectrum of fluctuations is already important in
the interaction propagator while vertex Cooperons can still be taken
at zero field. As the result $\delta\sigma^{a}_{MT}(H)$ turns out to
be logarithmic in $H$. At even higher fields $\omega_{H}\gg\ts^{-1}$
superconducting fluctuations are strongly suppressed and
corresponding correction decays inversely proportional to the third
power of magnetic field. Quantitatively we find following asymptotes
for these limits:
\begin{equation}\label{MT-a-H}
\frac{\delta\sigma^{a}_{MT}(H)}{\sigma_D}=\frac{1}{\pi^3g^2}\left\{
\begin{array}{lc}
(T_{c}\ts)^3\ln^{2}\!\!\frac{1}{\omega_{H}\ts} &
\tgl^{-1}\ll\omega_{H}\ll\ts^{-1}\,,\\ \frac{}{} \\
4.46(T_c/\omega_{H})^3 & \omega_{H}\gg\ts^{-1}\,.
\end{array}
\right.
\end{equation}
Similar analysis can be carried for Eqs.~\eqref{qq-sigma-MT-b} and
\eqref{qq-sigma-MTD}. We find that interference contribution
$\delta\sigma^{b}_{MT}(H)$ follows the same asymptotes as
$\delta\sigma^{a}_{MT}(H)$ while mixed term $\delta\sigma_{MTD}(H)$
is negative and differs from Eq.~\eqref{MT-a-H} only by the
numerical coefficient $2/3$ in the limit
$\tgl^{-1}\ll\omega_{H}\ll\ts^{-1}$, and $1.4$ in the limit
$\omega_H\gg\ts^{-1}$. The magnetic field dependence of the
conventional AL and MT contributions was recently discussed in
Ref.~\onlinecite{Magneto-cond}. Finally, Fig.~\ref{Fig-3}
schematically shows anomalous correction to the conductivity induced
by the interacting fluctuations in the whole range of magnetic
fields.

\textit{Regularization}. Without an extrinsic static pair-breaking
$\ts\to\infty$ physical origin of the divergent MT contribution
$\delta\sigma^{b}_{MT}$ and mixed terms $\delta\sigma_{MTD}$ comes
from the softness of the Cooperon. This is exactly the same problem
that exists for the conventional MT contribution Eq.~\eqref{MT} and
thus, regularization is achieved by following the prescription given
in the Ref.~\onlinecite{Reizer}, which allowed to regularize
$\delta\sigma_{MT}$ [Eq.~\eqref{MT-Reizer}]. The main idea is to
include Cooperon self-energy~\cite{Reizer,Brenig,SLF}
\begin{eqnarray}\label{self-energy}
\Sigma_{\epsilon,-\epsilon}=\frac{1}{\pi\nu}\sum_q\int d\omega
\left[\coth\frac{\omega}{2T}+\tanh\frac{\epsilon-\omega}{2T}\right]\nonumber\\
\times\mathrm{Im}[L^A_{q,\omega}]\mathrm{Re}[C^{R}_{q,2\epsilon-\omega}]
\end{eqnarray}
into the general scheme of calculations. It is important to realize
that this object is strongly $\epsilon$ dependent in the broad range
of energies~\cite{Reizer}
\begin{equation}
\Sigma_{\epsilon,-\epsilon}=\frac{T^2}{2\pi g|\epsilon|}\,,\quad
\tgl^{-1}\lesssim\epsilon\ll T\,.
\end{equation}
Since $\Sigma_{\epsilon,-\epsilon}$ enters now the Cooperon instead
of $\ts^{-1}$ an integration in Eq.~\eqref{sigma-MT-b} over the
fermionic energy $\epsilon$ must be completed carefully. An
inspection of the integrand reveals that $\epsilon\simeq T/\sqrt{g}$
give the most important contribution. After an explicit calculation
we find
\begin{equation}\label{MT-b-reg}
\frac{\delta\sigma^{b}_{MT}}{\sigma_D}\simeq
\frac{1}{4\sqrt{2}\pi^{3/2}}
\frac{1}{\sqrt{g}}\ln^{2}\left(\frac{\tgl}{\tphi}\right)\,,\quad
\tphi^{-1}\simeq\frac{T}{\sqrt{g}}\,,
\end{equation}
which is applicable in the temperature range
$Gi\lesssim\frac{T-T_c}{T_c}\lesssim\sqrt{Gi}$. Thus, inclusion of
$\Sigma_{\epsilon,-\epsilon}$ self-consistently generates an
intrinsic dephasing time $\tphi$ which regularizes
$\delta\sigma^{b}_{MT}$. Even more importantly, quantum dynamic
pair-breaking encoded by the Cooperon self-energy changes
dramatically the temperature dependence of this singular interaction
correction and leads to its saturation apparent from
Eq.~\eqref{MT-b-reg} [$\delta\sigma^{a}_{MT}$ also saturates in this
case as shown in Ref.~\onlinecite{Reizer}]. The same regularization
approach applied to the mixed term $\delta\sigma_{MTD}$ gives a
contribution that is three times smaller then $\delta\sigma_{MTD}$,
which concludes our analysis.

This work at ANL was supported by the U.S. Department of Energy
under Contract No.~DE-AC02-06CH11357.



\begin{thebibliography}{99}
\bibitem{note}
It is rather interesting that MT contributions remain to be
important even far away from $T_c$ and even in the case of repulsion
between electrons. Larkin [Ref.~\onlinecite{Larkin}] had shown that
conventional MT term [Eq.~\ref{sigma-MT}] partially compensates for
the contribution to conductivity due to weak-localization. This
observation significantly improved an agreement between the theory
and experiment. Due to its potential importance and for completeness
we also calculated $\delta\sigma^{a(b)}_{MT}$ away from the critical
region, $\ln\frac{T}{T_c}\gg1$, and found:
$$
\frac{\delta\sigma^{a(b)}_{MT}}{\sigma_D}\backsimeq\frac{c_{a(b)}}{g^2}
\frac{\ln^2[\lambda T\ts]}{\ln^4(T/T_c)}\,,\quad
\lambda=\min\{4\pi,\ln(T/T_c)\}\,,
$$
with coefficients $c_a=(13\pi^2+12)/36864$ and
$c_b=(\pi^2+12)/9216$.
\end{thebibliography}
\end{document}